\documentclass[10pt]{article}

\usepackage{amsmath}
\usepackage{amssymb}

\usepackage{graphicx}

\usepackage{cite}

\usepackage{color} 

\usepackage[labelfont=bf,labelsep=period,justification=raggedright]{caption}

\makeatletter
\renewcommand{\@biblabel}[1]{\quad#1.}
\makeatother

\date{}

\begin{document}

\begin{flushleft}
{\Large
\textbf{Study on Multicellular Systems Using a Phase Field Model}
}
\\
Makiko Nonomura$^{1,2,\ast}$
\\
\bf{1}
 Department of Mathematical Information Engineering, College of Industrial Technology, Nihon University,
 1-2-1 Izumicho, Narashino-shi, Chiba 275-8575, Japan.\\
\bf{2}
Japan Science and Technology Agency,  PRESTO,  4-1-8 Honcho, Kawaguchi-shi, Saitama 332-0012, Japan.
\\
$\ast$ E-mail: nonomura.makiko@nihon-u.ac.jp
\end{flushleft}

\section*{Abstract}

A model of multicellular systems with several types of cells is developed from the phase field model. The model is presented as a set of partial differential equations of the field variables, each of which expresses the shape of one cell. The dynamics of each cell is based on the criteria for minimizing the surface area and retaining a certain volume. The effects of cell adhesion and excluded volume are also taken into account. The proposed model can be used to find the position of the membrane and/or the cortex of each cell without the need to adopt extra variables. This model is suitable for numerical simulations of a system having a large number of cells.  The two-dimensional results of cell adhesion, rearrangement of a cell cluster, and chemotaxis as well as the three-dimensional results of cell clusters on the substrate are presented.

\section*{Introduction}

In order to investigate the structural patterns of cellular systems, several cell models have been reported, including the vertex dynamics model \cite{Honda2004,Nagai2001}, the center dynamics model \cite{Honda1978, Honda1984}, and the cellular Potts model \cite{Glazier1993,Graner1992}. 
Both the vertex dynamics model and the center dynamics model express cell patterns using polygons. 
In the vertex dynamics model, a cell or a cluster of cells is represented by a polygon formed by linking several vertices. Each vertex is driven by forces acting on it. This model has been adopted for morphogenesis in Xenopus notochords as well as cell deformation and rearrangement by applying mechanical forces \cite{Honda2004, Weliky1991}. In the center dynamics model, a node represents a cluster of cells and receives forces from its neighboring nodes. Cell aggregation, locomotion, rearrangement, and morphogenesis in vertebrate limb buds have been investigated using this model \cite{Graner1993, Honda1978, Honda1979, Honda1984, Morishita2008}. 
Although the mechanical processes during tissue developments can be well investigated, artificial treatments are required for numerical simulations in these models based on polygons. For example, in the vertex dynamics model, cell rearrangement is realized by manually exchanging two vertices that approach each other \cite{Honda2004}. In the center dynamics model, in order to express the cell division, it is necessary to add a new node in the vicinity of the existing node \cite{Honda1984, Morishita2008}. 

In contrast, the cellular Potts model represents each cell as a cluster of grid points under the constraint of constant volume. Thus, the artificial treatments mentioned above are not required for simulations in this model. We can investigate the deformation of an individual cell in a multicellular system using this model, considering the effects of excluded volumes and adhesions of the cells. This model successfully described several biological behaviors \cite{Newman2007}.  For example, numerical calculations with regard to cell sorting, biofilm formation, and chemotactic movement have been performed\cite{Glazier1993, Graner1992, Kafer2006, Poplawski2008}.  However, running the simulations requires fluctuations, and the forces between cells are not expressed directly in this model. 

Therefore, we consider a new type of a model for multicellular systems, which is based on the phase field model. The effects of cell adhesion and excluded volume are taken into account.  In the proposed model, the free energy is described in terms of a vector variable, the number of components of which is equivalent to the total number of cells in the system. The shape of one cell is expressed by one component of the vector variable. The time evolutions are described by a set of partial differential equations that are obtained by taking the functional derivative of the free energy.  Thus, fluctuations are not required for numerical simulations.   In addition, by adopting auxiliary variables that are used for calculation of the interactions between the cells, a program that consumes little computational memory can be designed. That is to say, the proposed model can be used to describe a system containing a large number of cells.
The proposed model differs from previous models of multicellular systems in that the position of the cell membrane and/or cortex can also be expressed without the need to adopt extra variables because the phase boundary interface is treated as a diffuse interface of finite width using the phase field method.  

The phase field model  has been applied to a wide range of problems, such as crystal growth \cite{Folch1999, Karma1998, Karma2001, Kobayashi1993, Kockelkoren2003}. 
Very recently, the cell shape of the fish keratocyte has been modeled using this method, where the membrane bending force and the surface tension of the cell were considered \cite{Shao2010}. However, to our knowledge, this is the first report applying the phase field method to the multicellular system. 

\section*{Results}

\subsection*{Model Equation}

We consider a multicellular system containing several types of cells and allow changes in the size and adhesive strength of each cell type. As a first step,  we express the shape of one cell using the phase field method. 

The following Ginzburg-Landau free energy is considered:
\begin{eqnarray}
E[u]=\int_\Omega \left[\frac{D_0}{2}|\nabla u|^2 + \frac{1}{4}u^2(1-u)^2\right]d\mathbf{r}+\frac{\alpha_0}{12}(V_0-v(u))^2 ,
\label{eq:GL2}
\end{eqnarray}
where $\Omega$ denotes the area of the system, and the coefficients $D_0$, $\alpha_0$, and $V_0$ are positive constants. 
The variable $u(\mathbf{r}, t)$ is an order parameter referred to as the phase field, where $\mathbf{r}$ is the position, and $t$ is the time. 
The function $v(u)$ is given as
\begin{eqnarray}
v(u)=\int_{\Omega} h(u) d\mathbf{r} ,
\label{eq:v}
\end{eqnarray}
where the function $h(u)$ is defined as 
\begin{eqnarray}
h(u)=u^2(3-2u).
\end{eqnarray}
By taking the functional derivative of Equation \ref{eq:GL2} with respect to $u$, the time evolution of $u$ is derived as follows: 
\begin{eqnarray}
\tau\frac{\partial u}{\partial t} &=& -\frac{\delta E}{\delta u}
\nonumber\\
&=&D_0 \nabla^2 u + u(1-u)\left(u-\frac{1}{2}+f_0(u)\right) ,
\label{eq:AC2}\\
f_0(u) &=& \alpha_0(V_0-v(u)) ,
\label{eq:f_0}
\end{eqnarray}
where $\tau$ is a positive constant. Equation \ref{eq:AC2} guarantees the monotonic decrease in the free energy. 

If the function $f_0$ is set to be constant $\bar f$, then Equation \ref{eq:AC2} is referred to as the Allen--Cahn equation in the field of materials science and is known for having a smooth front solution connecting the regions $u = 1$ and $u = 0$. The Allen--Cahn equation can easily be solved in one dimension as $u=\{1-\tanh[(x-\mathcal{V} t)/(2\sqrt{2D_0})]\}/2$, where the front velocity $\mathcal{V}=\sqrt{2D_0} \bar f/\tau$. This means that the front moves such that the region of $u = 1$ ($u = 0$) expands if $\bar f>0$ ($\bar f<0$). 

Note that the function $v(u)$ can be regarded as the volume of the region in which $u=1$ because $h(1) = 1$ and $h(0) = 0$. Therefore, as discussed above, Equation \ref{eq:f_0} indicates that the region of $u=1$ expands (shrinks) until $v(u)=V_0$ when $v(u)<V_0$ ($v(u)>V_0$). The last term of Equation \ref{eq:GL2}, which has a minimum at $v(u)=V_0$, also expresses the constraint of the constant volume of $u$.  

\begin{figure}[htbp]
\begin{center}
\includegraphics[scale=1.0, angle=0]{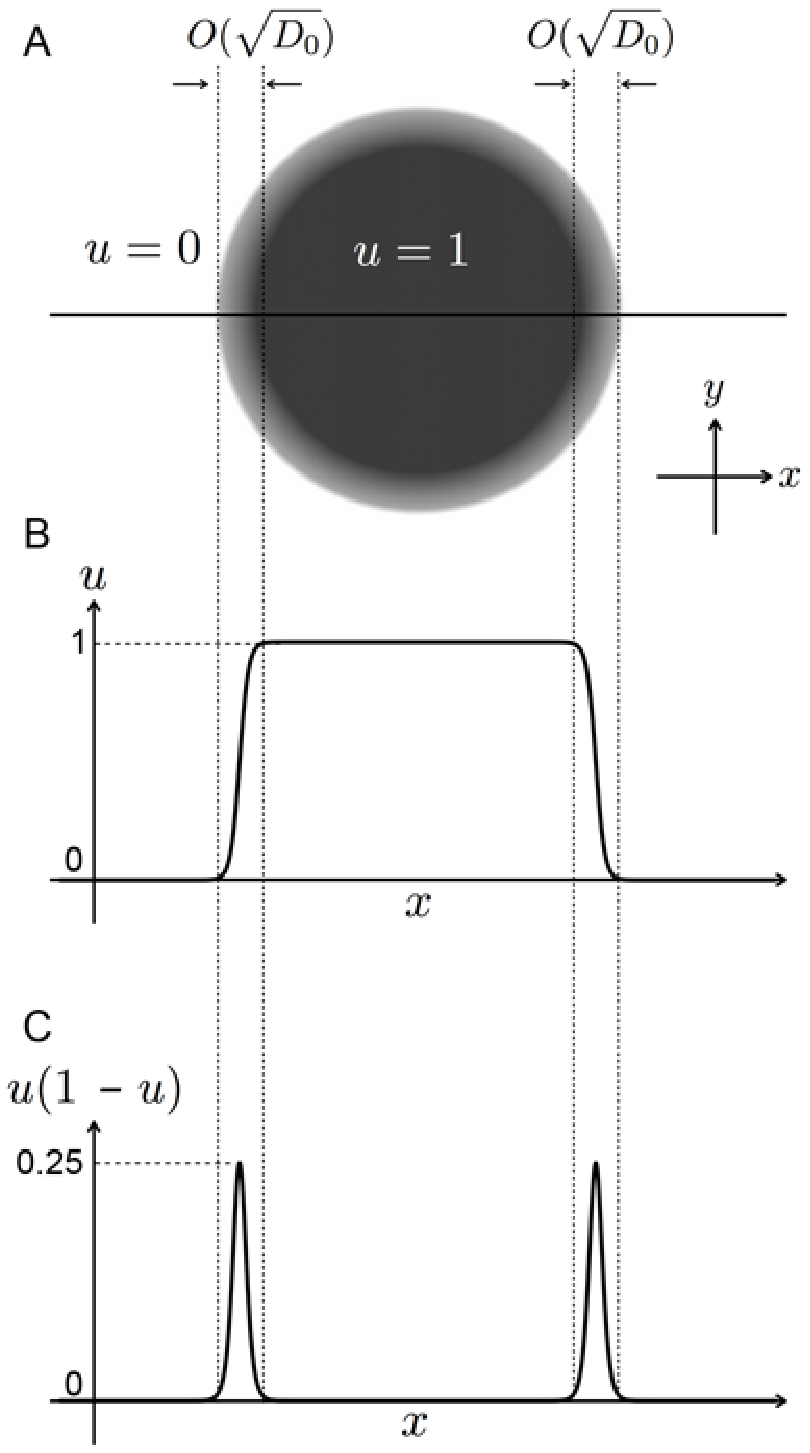}
\caption{
{\bf Shape of the phase field $u$.} The integral of  $u$ over $\mathbf{r}$ is set to be maintained. Panel A: top view. Panels B and C: profiles of $u$ and $u(1-u)$ at the centerline in Panel A, respectively.
}
\label{fig:u}
\end{center}
\end{figure}

As shown in Figure \ref{fig:u}, the region of $u =1$ takes the form of a circle in two dimensions and a sphere in three dimensions in the steady state.  Thus, the shape of the cell in the simplest case can be described by a single-order parameter $u$, such that $u\ge u_{cell}$ in the region with the cell ($<u_{cell}$ in the region not taken up by the cell) with a constant $u_{cell}\in(0,1)$. Based on the fact that $u$ has an interface with a thickness on the order of $\sqrt{D_0}$, the cell cortex can also be expressed as a function of $u$, e.g., $u(1-u)$ (see Figure \ref{fig:u}C). 

\begin{figure}[htbp]
\begin{center}
\includegraphics[scale=1.0, angle=0]{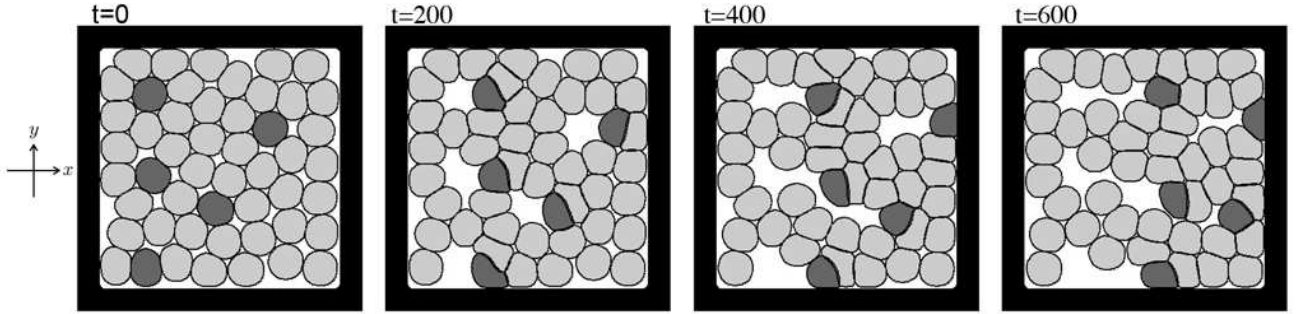}
\caption{
{\bf Two-dimensional results of chemotactic movement of cells.} The case of fifty cells ($M=50$) of two types ($L=2$) is considered. Light gray (dark gray) areas indicate the region of $u_m \ge 0.2$, for the case in which the $m$-th cell is a type-1 (type-2) cell.  Black areas represent the walls ($s \ge 0.5$). Numerical calculation was performed with  $\mu(1) = 0.0$ and $\mu(2) = 1.0$. The other parameters are set as follows: size of the simulation box $\Omega=10\times 10$, size of the spatial grid $\delta=0.05$, time increment $dt =0.01$, $\tau_u= 1$, $D_u(1)=D_u(2)=0.001$, $V(1)=V(2)=1$, $\alpha(1)=\alpha(2)=1$,  $\beta(1,1)=\beta(1,2)=\beta(2,2)=\beta_s(1)=\beta_s(2)=1$, $\gamma(1)=\gamma(2)=0$,  $\eta(1,1)=\eta(1,2)=\eta(2,2)=\eta_s(1)=\eta_s(2)=0$, $\epsilon_w=2\sqrt{0.002}$, $c_0=0.01$, $x_l=y_b=0.8$, and $x_r = y_t=9.2$.}
\label{fig:chem}
\end{center}
\end{figure}

In order to describe the multicellular system, a vector variable $\mathbf{u}(\mathbf{r}, t)=(u_1(\mathbf{r}, t),  \cdots, u_{M}(\mathbf{r}, t))$ is considered, where $M$ is the total number of cells in the system. The component $u_m(\mathbf{r}, t)$ ($m=1,\cdots, M$) describes the shape of the $m$-th cell. We also use the variable $s(\mathbf{r}, t)$ to represent the shape of substances interacting with the cells, such as the wall (Figures \ref{fig:hipparu} and \ref{fig:chem}), the substrate (Figures \ref{fig:3D}), and the ECM. 

The model free energy for  the multicellular system is written as
\begin{eqnarray} 
E[\mathbf{u}, s]=E_{cell}[\mathbf{u}]+E_{int}[\mathbf{u}]+E_{s}[\mathbf{u}, s] ,
\label{eq:enetot}
\end{eqnarray}
where $E_{cell}$ determines the shape of the cell, $E_{int}$ describes the interactions between each cell, and $E_{s}$ expresses the interactions between the cells and substances external to them. The form of $E_{cell}$ is obtained by modifying Equation \ref{eq:GL2} using the vector variable as follows:
\begin{eqnarray} 
E_{cell}[\mathbf{u}] &=& \sum_m
\int_\Omega\left[
\frac{D(\ell_m)}{2}|\nabla u_m|^2
+\frac{1}{4}u_m^2(1-u_m)^2
\right] d\mathbf{r}
\nonumber\\
&+&\sum_m \frac{\alpha(\ell_m)}{12} (V(\ell_m)-v(u_m))^2 ,
\label{eq:enecell}
\end{eqnarray}
where $\ell_m$ is the cell type of the $m$-th cell. The coefficients $D(\ell)$, $\alpha(\ell)$, and $V(\ell)$ ($\ell=1,\cdots L$) are positive constants, where $L$ is the total number of cell types in the system. As discussed in the paragraph below Equation \ref{eq:f_0}, Equation \ref{eq:enecell} indicates that the thickness of the cell interface is on the order of $\sqrt{D(\ell)}$ and that the speed at which the volumes of the type-$\ell$ cells approach the target volume $V(\ell)$ is controlled by the value of $\alpha(\ell)$. That means $\alpha(\ell)$ determines the cell size growth.
Here, $E_{int}$ can be presented in the following form:
\begin{eqnarray} 
E_{int}[\mathbf{u}] &=& 
\sum_{m}\sum_{m'\ne m}\frac{\beta(\ell_m,\ell_{m'})}{6}\int_\Omega h(u_m) h(u_{m'}) d\mathbf{r}
\nonumber\\
&+&\sum_{m}\sum_{m'\ne m}\frac{\eta(\ell_m,\ell_{m'})}{6} \int_\Omega   \nabla h(u_m)\cdot \nabla h(u_{m'}) d\mathbf{r}
\nonumber\\
&+&\sum_{m}\frac{\gamma(\ell_m)}{12} \int_\Omega   |\nabla h(u_m)|^2 d\mathbf{r} ,
\label{eq:eint}
\end{eqnarray}
where $\beta(\ell, \ell')$, $\eta(\ell,\ell')$, and $\gamma(\ell)$ ($\ell, \ell' =1,\cdots L$) are positive constants. The first term on the right-hand side of Equation \ref{eq:eint} represents the effect of the excluded volume by increasing the energy if the cells overlap, whereas the second term represents the effect of cell adhesion by decreasing the energy if the cell cortices overlap. This adhesion term becomes negative in the region in which cell adhesion occurs. In order to prevent divergence due to this adhesion term, we introduce the third term on the right-hand side of Equation \ref{eq:eint} with the condition whereby $\gamma(\ell)>\eta(\ell,\ell)$. Similarly, the interaction between cells and substances external to the cells is expressed as follows:
\begin{eqnarray} 
E_{s}[\mathbf{u}, s] &=& 
\sum_{m}\frac{\beta_s(\ell_m)}{6}\int_\Omega h(u_m)h(s) d\mathbf{r}
\nonumber\\
&+&\sum_{m}\frac{\eta_s(\ell_m)}{6} \int_\Omega   \nabla h(u_m)\cdot \nabla h(s) d\mathbf{r} ,
\end{eqnarray}	
where $\beta_s(\ell)$ and $\eta_s(\ell)$ ($\ell=1,\cdots, L$) are positive constants.

\begin{figure}[htbp]
\begin{center}
\includegraphics[scale=1.0, angle=0]{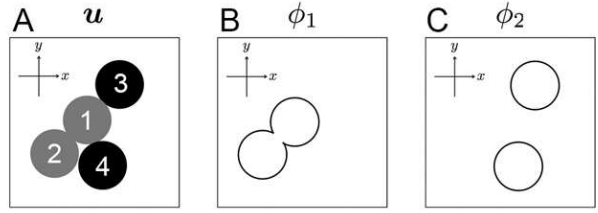}
\caption{
{\bf Schematic diagram of $\mathbf{u}$ and $\mathbf{\phi}$.} In Panel A, type-1 ($m=1$ and $2$) and type-2 ($m=3$ and $4$) cells are represented by gray and black circles, respectively. The contours of $\phi_1$ and $\phi_2$ are indicated by curved lines in Panels B and C, respectively.
}
\label{fig:fig-psi}
\end{center}
\end{figure}

Taking the functional derivative of Equation \ref{eq:enetot} with respect to $u_m$, the following time evolution equations are obtained:
\begin{eqnarray} 
\tau_u\frac{\partial u_m }{\partial t}
& =&D(\ell_m)\nabla^2 u_m
\nonumber\\
 &+& u_m (1-u_m)\left(u_m-\frac{1}{2}+f(u_m, s, \mathbf{\phi})\right) 
 \nonumber\\
 &+&g_{int}(u_m,\mathbf{\phi})+g_{s}(u_m,s) ,
 \label{eq:u}\\[1ex]
f(u_m, s, \mathbf{\phi}) &= &\alpha(\ell_m)(V(\ell_m)-v(u_m))
\nonumber\\
&-&\sum_\ell\beta(\ell_m,\ell)\left[  \phi_\ell -h(u_m)\delta_{\ell_m, \ell}\right] 
-\beta_s(\ell_m)h(s) ,
\label{eq:f_u}\\
g_{int}(u_m,\mathbf{\phi})&= &
\sum_{\ell} \eta(\ell_m, \ell) \nabla \left[ u_m (1-u_m)\nabla \{\phi_\ell-h(u_m)\delta_{\ell_m,\ell}\}\right]
\nonumber\\
&+&\gamma(\ell_m)\nabla[u_m(1-u_m)\nabla h(u_m)] ,
\label{eq:g_adh}\\
g_{s}(u_m,s)&=&
 \eta_s(\ell_m) \nabla \left[ u_m (1-u_m)\nabla h(s)\right] ,
\label{eq:g_s}
\end{eqnarray}
where $\tau_u$ is a positive constant, and $\delta_{i,j}$ is the Kronecker delta, which is $\delta_{i,j} =1$ ($\delta_{i,j}=0$) if $i=j$ ($i\ne j$). The vector variable $\mathbf{\phi}(\mathbf{r}, t)=(\phi_1(\mathbf{r}, t), \cdots, \phi_L(\mathbf{r}, t))$ is an auxiliary variable that is defined as follows:
\begin{eqnarray}
\phi_\ell(\mathbf{r}, t) = \sum_m h(u_m(\mathbf{r}, t))\delta_{\ell_m,\ell} .
\end{eqnarray}
As shown in Figure \ref{fig:fig-psi}, the region occupied by the type-$\ell$ cells can be identified by $\phi_\ell$.

Note that the interaction terms in Equation \ref{eq:u} are not written explicitly in terms of the variables $u_{m'}$ $(m'=1,\cdots, M \ne m)$ but are instead written in terms of the auxiliary variable $\mathbf{\phi}$. 
Moreover, the components $E_{int}$ and $E_{s}$ can also be presented in terms of $\mathbf{\phi}$, as follows:
\begin{eqnarray}
E_{int}[\mathbf{u}]& =& 
\sum_\ell\sum_{\ell'}\frac{\beta(\ell,\ell')}{12}\int_\Omega \phi_\ell\phi_{\ell'} d\mathbf{r}
\nonumber\\
&-&\sum_{m}\frac{\beta(\ell_m,\ell_m)}{12}\int_\Omega h(u_m)^2 d\mathbf{r}
\nonumber\\
&+&\sum_{\ell}\sum_{\ell'}\frac{\eta(\ell,\ell')}{12} \int_\Omega   \nabla \phi_{\ell}\cdot \nabla \phi_{\ell'} d\mathbf{r}
\nonumber\\
&+&\sum_{m} \frac{\gamma(\ell_m)-\eta(\ell_m,\ell_m)}{12}\int_\Omega |\nabla h(u_m)|^2 d\mathbf{r} ,
\label{eq:Eint2}\\
E_{s}[\mathbf{u}, s] &=& 
\sum_{\ell}\frac{\beta_s(\ell)}{6}\int_\Omega \phi_{\ell}h(s) d\mathbf{r}
\nonumber\\
&+&\sum_{\ell}\frac{\eta_s(\ell)}{6} \int_\Omega   \nabla \phi_{\ell}\cdot \nabla h(s) d\mathbf{r} .
\end{eqnarray}
We adopted the second term on the right-hand side of Equation \ref{eq:f_u} and the first term on the right-hand side of Equation \ref{eq:g_adh} to express the excluded volumes and the cell adhesions, respectively, because these terms are the simplest among the several alternatives, which can be written in terms of $\mathbf{\phi}$ both in the time evolution equation for $\mathbf{u}$ and the component $E_{int}$.

\subsection*{Numerical implementation}

In order to rapidly simulate a system having numerous cells, it is important to design a program that does not consume a large amount of computational memory and to increase the simulation speed. These two requirements are easily satisfied because Equation \ref{eq:u} is not written explicitly in terms of $u_{m'}$ ($m'=1,\cdots, M \ne m$). Once $\mathbf{\phi}$ is obtained for each time step, the time evolution of $u_m$ can be computed independent of $u_{m'}$. Such a program is fully compatible with parallel computation. Moreover, the shape of the $m$-th cell can be obtained by computing the equation for $u_m$ within the small region $\Omega_m$, which covers the region of $u_m>0$. This reduces the computational memory and increases the simulation speed.
The position $\mathbf{r}_m(t)$, which indicates the center position of $\Omega_m$ measured for the entire system, must be moved along with the movement of the center position of the $m$-th cell. Since $u_m=0$ is realized outside the region $\Omega_m$, the Dirichlet boundary condition must always be set for the small region $\Omega_m$. 

The estimation of the required memory is described below. Since the number of cell types $L$ is generally much smaller than the number of cells $M$, the memory increase by introducing $\mathbf{\phi}$ becomes smaller than the memory decrease by computing $u_m$ within the small region $\Omega_m$. For simplicity, we assume that each of the cells has the same volume, i.e., $V(1)=\cdots=V(L)=V$ and that the entire system is covered by the cells, i.e., $\Omega\sim VM$. Then, the computational memories for $\mathbf{u}$, $\mathbf{r}(t)=(\mathbf{r}_1(t),\cdots, \mathbf{r}_M(t))$, and $\mathbf{\phi}$ are roughly estimated as $VM/\delta^d$, $d M$, and $LVM/\delta^d$, respectively, where $d$ is the spatial dimension and $\delta$ is the size of the spatial grid. Therefore, the total memory required to compute Equation \ref{eq:u} using $\mathbf{\phi}$ is linearly dependent on $M$. On the other hand, in order to compute cell-cell interactions without using $\mathbf{\phi}$, the value of $\mathbf{u}$ must be preserved over the entire region $\Omega$. Then, the computational memories for solving Equation \ref{eq:u} increase by $VM/\delta^d \times M \propto M^2$. These estimations reveal that the introduction of $\mathbf{\phi}$ is very useful for computation in the case of a system that contains a large number of cells, even in three dimensions.

\subsection*{Numerical Simulation}

\begin{figure}[htbp]
\begin{center}
\includegraphics[scale=1.0, angle=0]{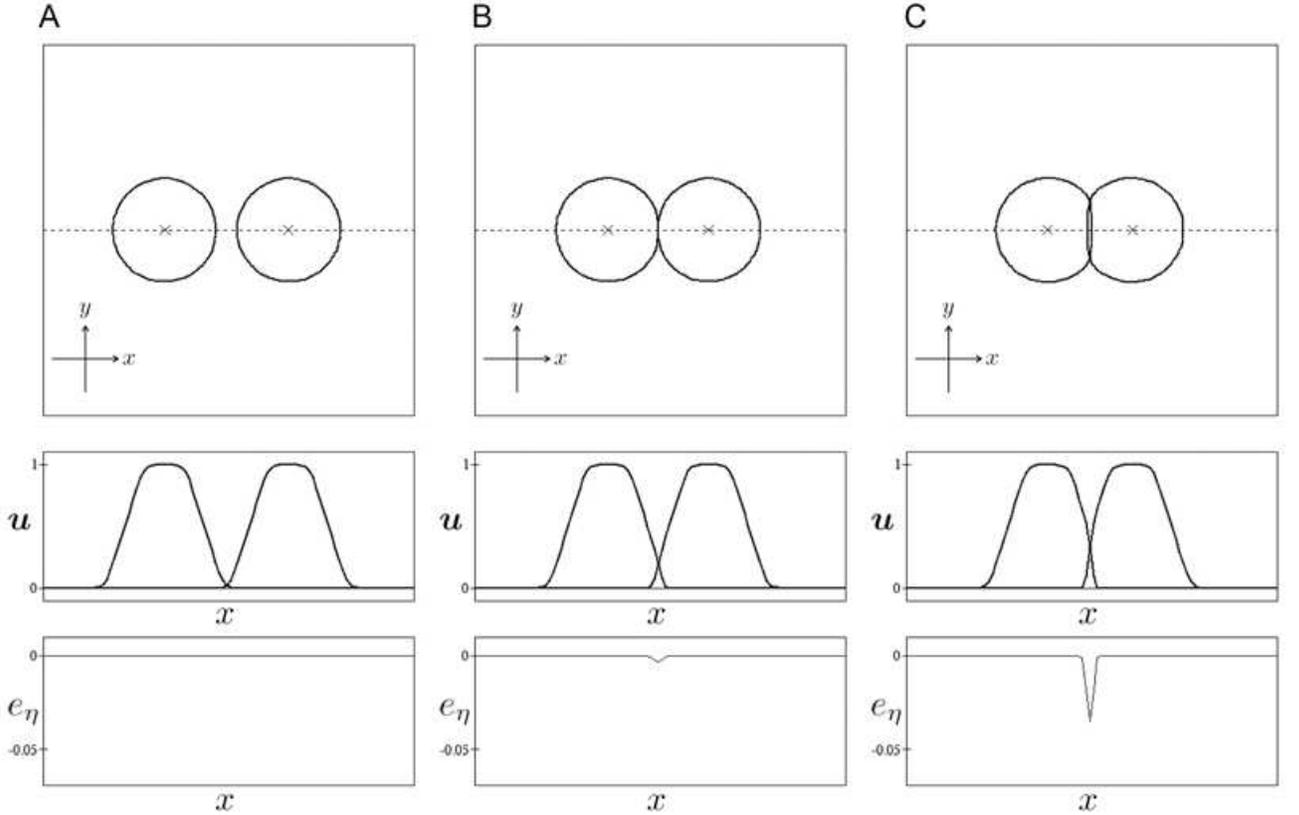}
\caption{
{\bf Two-dimensional results of cell adhesions.} The case of two cells ($M=2$) of the same type ($L=1$) is considered. 
Numerical calculations were performed with $\eta(1,1)=0.000$ in Panel A, $\eta(1,1)=0.004$ in Panel B, and $\eta(1,1)=0.008$ in Panel C.
The top row shows contour plots of $u_m = 0.2$ ($m = 1, 2$). The $\times$ symbol indicates the centers of cells. The middle and bottom rows show the profiles of $u_m$ and $e_\eta$ along the dotted line shown in the top row. The size of the simulation box is $\Omega = 5 \times 5$, and the size of the spatial grid is $\delta=0.05$. The time increment is $dt=0.01$. The remaining parameters are set as follows: $\tau_u = 1$, $D(1)=0.001$, $V(1)=1$, $\alpha(1)=1$, $\gamma(1)=0.01$, $\beta(1,1)=1$, and $\beta_s(1)=\eta_s(1)=0$.
}
\label{fig:adhesion}
\end{center}
\end{figure}

Figure \ref{fig:adhesion} shows the numerical results for two cells of the same type, i.e., $M=2$ and $\ell_1=\ell_2 =L =1$, with different adhesion strengths. The curves in the top row of the graphs indicate the contour lines of $u_m = 0.2$ ($m=1$ and $2$), and the $\times$ symbols indicate the positions of the centers of the cells $ \mathbf{r}_m = (\int_\Omega \mathbf{r} u_m d \mathbf{r})/(\int_\Omega u_m d \mathbf{r})$. 
The variable $e_\eta(\mathbf{r},t)$ is given as follows:
\begin{eqnarray}
e_\eta(\mathbf{r},t)=\displaystyle\sum_m\sum_{m'\ne m}\frac{\eta(\ell_m,\ell_{m'})}{6}\nabla h(u_m(\mathbf{r}, t))\cdot\nabla h(u_{m'}(\mathbf{r}, t)) .
\label{eq:eeta}
\end{eqnarray}
The integral over $\mathbf{r}$ of $e_\eta$ is identical to the second term on the right-hand side of Equation \ref{eq:eint}.
The $u_m$ and $e_\eta$ profiles along the dotted line in the top row have been plotted in the middle and bottom rows of the graphs, respectively. Since $e_\eta$ has a non-zero value only in regions in which cell adherence occurs, $e_\eta$ is an indicator of locations at which cell adherence occurs. Initially, the distance between the centers of cells is set to $1.6000$. After a sufficiently long simulation time ($t = 50,000$), the two cells move closer to each other as the value of $\eta(1,1)$ increases, such that the distances between the cell centers are $1.662$ in the case of Panel A with $\eta(1,1)=0.0000$, $1.355$ in the case of Panel B with $\eta(1,1)=0.004$, and  $1.156$ in the case of Panel C with $\eta(1,1)=0.008$. 

Figure \ref{fig:3D} shows snapshots of three-dimensional simulations at $t=500$. The periodic boundary conditions are imposed on $\Omega$. The solid substrate is introduced by setting the variable $s$ as $s(\mathbf{r})=(1-\tanh((z-z_f)/\epsilon_f))/2$, where $z_f$ and $\epsilon_f$ are positive constants. The light gray surfaces are contour plots of $u_m = 0.1$ ($m=1,\cdots, 10$) and the dark gray surfaces represent contour plots of $s= 0.1$.  We set  $\eta(1,1)$ as $0.0000$, $0.0100$, and $0.0219$ for the simulations shown in Panels A, B, and C, respectively, where the other parameters are the same for all cases. If the cell adhesions are weak, the cells push against each other, and their positions are determined as shown in Panel A. On the other hand, for the case in which the cell adhesions are sufficiently strong, the cell positions are decided by the pulling force between cells, and the surface of the cell layer becomes flat, as shown in Panel C.

\begin{figure}[htbp]
\begin{center}
\includegraphics[scale=1.0, angle=0]{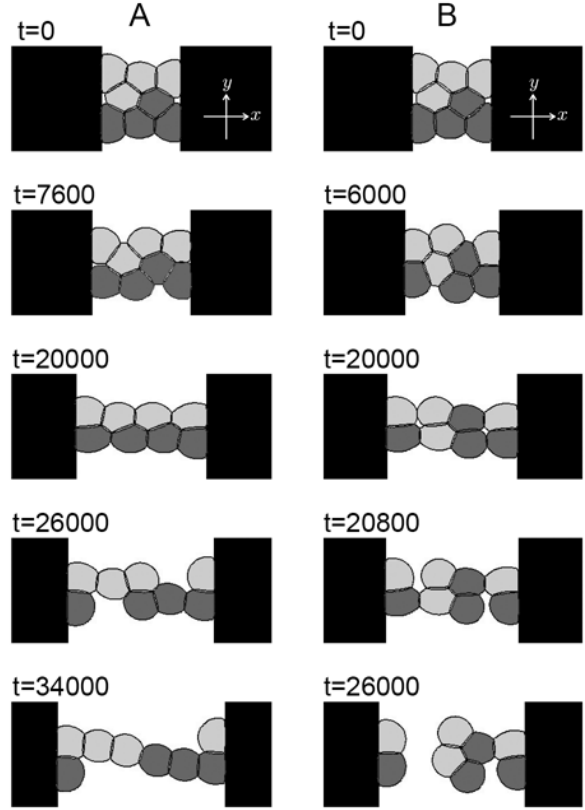}
\caption{
{\bf Two-dimensional results of cell deformation and rearrangement in a cluster.} The cluster is composed of eight cells ($M=8$) of two types ($L=2$). Light and dark gray areas represent the region of $u_m\ge 0.2$. Light gray areas indicate the locations of type-1 cells, and dark gray areas indicate the locations of type-2 cells. Black areas represent the walls ($s\ge 0.5$). Numerical calculations were performed with $\eta(1,1)=\eta(2,2)=0.008$ and $\eta(1,2)=0.005$ in Panel A and $\eta(1,1)=\eta(2,2)=0.005$ and $\eta(1,2)=0.008$ in Panel B. The left and right walls are assumed to move at a uniform velocity, $x_l =7-\mathcal{V}_s t$, $x_r=13+\mathcal{V}_s t$, and $\mathcal{V}_s=0.0001$. The size of the simulation box is $\Omega = 20\times 15$, and the size of the spatial grid is $\delta=0.05$. The time increment is $dt=0.01$. The remaining parameters are set as follows: $\tau_u = 1$, $D_u(1)=D_u(2)=0.001$, $V(1)=V(2)=4$, $\alpha(1)=\alpha(2)=10$,
$\gamma(1)=\gamma(2)=0.01$, $\beta(1,1)=\beta(1,2)=\beta(2,2)=\beta_s(1)=\beta_s(2)=0.1$, 
$\eta_s(1)=\eta_s(2)=0.01$, and $\epsilon_s = 2\sqrt{0.002}$ .
}
\label{fig:hipparu}
\end{center}
\end{figure}

\begin{figure}[htbp]
\begin{center}
\includegraphics[scale=1.0, angle=0]{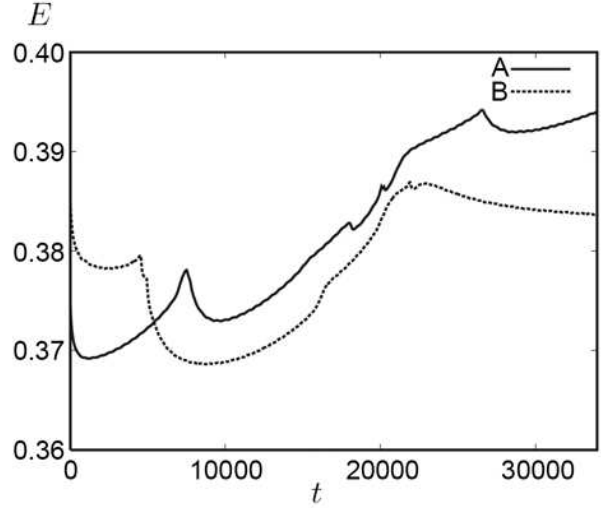}
\caption{
{\bf Plots of the total energy $E$ with respect to time.} The solid line shows the results for Figure \ref{fig:hipparu}A, and the dotted line shows the results for Figure \ref{fig:hipparu}B.
}
\label{fig:ene}
\end{center}
\end{figure}

Figure \ref{fig:hipparu} shows the numerical results for cell deformation and rearrangement. A cell cluster of $M = 8$ and $L = 2$ is sandwiched between two walls that move at a constant speed. In this calculation, considering the variable $s$ as an order parameter that corresponds to the walls, the time evolution of $s$ is calculated as 
$s(\mathbf{r}, t)=1-\left(1+\tanh((x-x_l(t))/\epsilon_s)\right)\left(1-\tanh((x-x_r(t))/\epsilon_s)\right)/4$, 
where $\epsilon_s$ is a positive constant. The locations of the left and right walls are denoted as $x_l(t)$ and $x_r(t)$, respectively. Panel A shows the results for the case in which the adhesion strength between cells of the same type is stronger than that between cells of different types ($\eta (1, 1) = \eta(2, 2) = 0.008$ and $\eta(1, 2) = 0.005$), whereas Panel B shows the results for the opposite case ($\eta (1, 1) = \eta(2, 2) = 0.005$ and $\eta(1, 2) = 0.008$). Light gray, dark gray, and black areas represent the positions of the type-1 cells, the type-2 cells, and the walls, respectively. Cells adhering to the walls are stretched by the moving walls, causing the cells to be deformed and rearranged. Cells that are rearranged as weakly adhered cells detach first. In Panel B, the cells separate into two groups at approximately $t = 22,000$ and relax to almost their original shape at $t = 26,000$. 
The time evolution of the total energy $E$ is plotted in Figure \ref{fig:ene}. The solid line shows the results for Panel A of Figure \ref{fig:hipparu}, and the dotted line shows the results for Panel B of Figure \ref{fig:hipparu}. There is no monotonic decrease in total energy because the walls stretch the cell clusters. Comparison of Figures \ref{fig:hipparu} and \ref{fig:ene} reveals that the energy decreases significantly when cell rearrangement occurs.

Finally, we show that the additional cell behavior can also be incorporated into the proposed model. For example, the chemotactic movement of the cell can be described by adding a new term, such as $g_{chem}=-\mu(\ell_m) \nabla\cdot (u_m\nabla c)$ to the right-hand side of Equation \ref{eq:u}, where the variable $c(\mathbf{r}, t)$ is the chemical concentration in extracellular regions. The parameter $\mu(\ell_m)$ indicates the sensitivity of the $m$-th cell to the gradient of $c$. Figure \ref{fig:chem} shows the time evolution of a system with cells having chemotaxis. Light gray and dark gray represent type-1 and type-2 cells, respectively. In this case, we  consider the variable $s$ as an order parameter that corresponds to the wall. The fifty cells are surrounded by the unmoving wall defined as $s(\mathbf{r})=2-(1+\tanh((x-x_l)/\epsilon_w))(1-\tanh((x-x_r)/\epsilon_w))/4-(1+\tanh((y-y_b)/\epsilon_w))(1-\tanh((y-y_t)/\epsilon_w))/4$, where $x_l$, $x_r$, $y_b$, $y_t$, and $\epsilon_w$ are positive constants. Cell adhesion is not considered in this simulation. By setting $\mu(1)=0.0$ and $\mu(2)=1.0$, it is assumed that the only type-2 cells can sense the gradient of the chemical concentration $c$. For simplicity, the form of $c$ is assumed not to be affected by $u_m$ or $t$ and is taken as $c=c_0 x$, where $c_0$ is a constant. It is found numerically that type-2 cells move toward the $c$-rich region by pressing against type-1 cells.

\section*{Discussion}

We proposed a new type of cell model based on a phase field model, including the effects of excluded volumes and cell adhesions. The proposed model is based on a concept similar to the cellular Potts model, but the time evolutions of cell shapes in the proposed model differ from those in the cellular Potts model. In the cellular Potts model, the time evolutions of the spins are computed by the Monte Carlo method, and thus the fluctuations are required for the time evolution. On the other hand, the time evolution equations in the present model are written in the form of partial differential equations, whereby fluctuations are not necessary in order to run the simulations. In addition, the proposed model is thought to be more appropriate for investigating problems in which a small volume variant must be accounted for, because the proposed model is continuous in any parameter. 

\begin{figure}[htbp]
\begin{center}
\includegraphics[scale=1.0, angle=0]{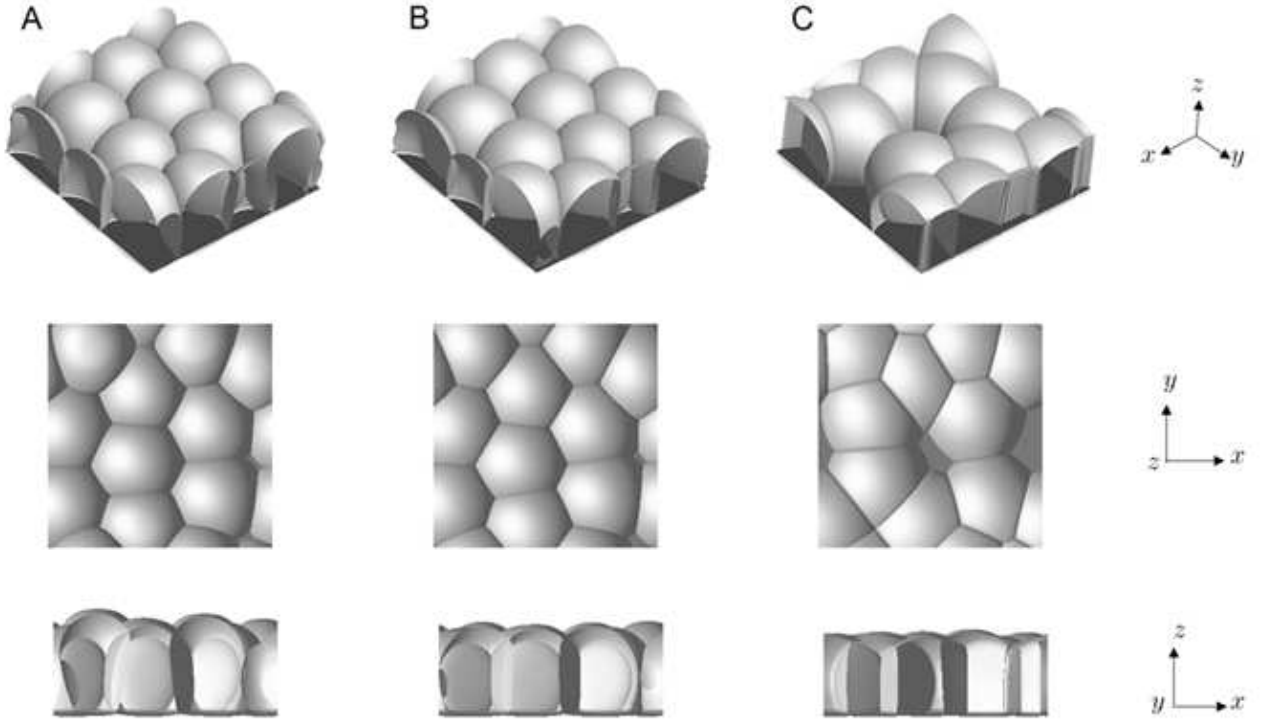}
\caption{
{\bf Three-dimensional results of cell adhesions on the substrate.} The case of 10 cells ($M=10$) of the same type ($L=1$) is considered. Numerical calculations were performed with $\eta(1,1)=0.0000$ in Panel A, $\eta(1,1)=0.0100$ in Panel B, and $\eta(1,1)=0.0219$ in Panel C.
Light and dark gray surfaces are contour plots of $u_m=0.1$ ($m=1,\cdots, 10$) and $s=0.1$, respectively. The diagonal, top, and side views for each result are shown in the top, middle, and bottom rows, respectively. The size of the simulation box is $\Omega = 5 \times 5 \times 4$, and the size of the spatial grid is $\delta=0.05$. The time increment is $dt=0.01$. The remaining parameters are set as follows: $\tau_u = 1$, $D(1)=0.001$, $V(1)=2.26$, $\alpha(1)=100$, $\gamma(1)=0.022$, $\beta(1,1)=\beta_s(1)=1$, $\eta_s(1)=0.01$, $z_f=0.8$, and $\epsilon_f=2\sqrt{0.001}$.
}
\label{fig:3D}
\end{center}
\end{figure}

Since the cell shapes are represented by interfaces of finite thickness, the proposed model has the potential to be applied to the investigation of not only shape changes due to interactions between cells (Figures \ref{fig:adhesion} and \ref{fig:3D}) and rearrangements of cells in clusters (Figure \ref{fig:hipparu}) but also phenomena requiring knowledge of the position of the cell membrane and/or cortex. It is easy to incorporate additional cell behaviors such as chemotaxis (Figure \ref{fig:chem}) into the proposed model by adding corresponding terms. At the stage of numerical implementation, a program that is suitable for parallel computing and that consumes little computational memory can be designed by introducing the auxiliary variable $\mathbf{\phi}$, which is commonly used for the calculation of interactions between cells. Therefore, simulations can be performed for a system with numerous cells, even in three dimensions (Figure \ref{fig:3D}). 

The proposed model can express the time evolution of changes in cell shape due to the interactions between cells, cell differentiation by changing the cell type, cell size growth, cell movement, and cell death by deleting the corresponding component of $\mathbf{u}$. Thus, this model may well provide a useful tool for approaching the problem of morphogenesis, although this remains a subject for future study in order to estimate the parameters by comparison with earlier models and with experimental data. We plan to include the cell division, in the process of which the cortex of the dividing cell is known to be important \cite{Akiyama2010, Grill2003, Thery2007}, in the present model and to approach the problem of morphogenesis.

\section*{Acknowledgments}
The authors would like to thank M. Akiyama, H. Kitahata, R. Kobayashi,  T. Sakurai,  T. Shibata,  and A. Tero for the valuable discussions. 

\bibliography{cell}

\section*{Figure Legends}

\end{document}